\documentclass[journal=apchd5,manuscript=article]{achemso}
\setkeys{acs}{usetitle = true}
\usepackage[version=3]{mhchem}
\usepackage{graphicx}
\graphicspath{{./}{./pics/}}
\usepackage{float}
\usepackage{color}
\usepackage{bm}
\usepackage{amsmath}
\usepackage{epstopdf}
\newcommand\pictc[5]{\begin{figure}[tp]
                       \centerline{\vspace{0mm}
\includegraphics[width=#1\columnwidth,height=0.5\textheight,keepaspectratio]{#3}}
                       \protect\caption{\protect\label{fig:#4} #5}\vspace{0mm}
                    \end{figure}            }
\newcommand\pict[4][1]{\pictc{#1}{!tb}{#2}{#3}{#4}}
\newcommand\rpict[1]{\ref{fig:#1}}

\newcounter{Fig}

\newcommand{\be}{\begin{equation}}
\newcommand{\ee}{\end{equation}}

\author{Wei Liu}
\email{wei.liu.pku@gmail.com}
\affiliation{College of Optoelectronic Science and Engineering, National University of Defense
Technology, Changsha, Hunan 410073, P. R. China}
\author{Andrey E. Miroshnichenko}
\affiliation{School of Engineering and Information Technology,
University of New South Wales, Canberra, ACT 2600, Australia}

\title{Beam steering with dielectric metalattices}
\SectionNumbersOff
\begin{document}


\newpage
\begin{abstract}
\textbf{We study optical wave manipulations through high-index dielectric metalattices in both diffractionless metasurface and diffractive metagrating regimes. It is shown that the collective lattice couplings can be employed to tune the excitation efficiencies of all electric and magnetic multipoles of various orders supported by each particle within the metalattice. The interferences of those adjusted multipoles lead to highly asymmetric angular scattering patterns that are totally different from those of isolated particles, which subsequently enables flexible beam manipulations, including perfect reflection, perfect transmission and efficient large-angle beam steering. The revealed functioning mechanism of manipulated interplays between lattice couplings and multipolar interferences can shed a new light on both photonic branches of metasurfaces and metagratings, which can potentially inspire many advanced applications related to optical beam controls.}
\end{abstract}

{\bf Keywords:} Beam steering; dielectric metalattices; multipolar interferences; high-index nanoparticles; Mie resonances.
\\\\
 The recent emergence and rapid developments of metasurfaces have led to a number of applications including but not limited to wavefront shaping and beam control, nano-imaging, highly efficient holograms, photovoltaics  and quantum interferences in both  linear and nonlinear regimes~\cite{HOLLOWAY_IEEEAntennasPropag.Mag._overview_2012,YU_NatMater_flat_2014,brongersma2014light,estakhri_recent_2016,zhang_advances_2016,glybovski_metasurface_2016,CHEN_Rep.Prog.Phys._review_2016,
GENEVET_Optica_recent_2017,DING_Rep.Prog.Phys._gradient_2017,SMIRNOVA_Optica_multipolar_2016,LI_Nat.Rev.Mater._nonlinear_2017,krasnok_nonlinear_2017}. Though controversial about its origin~\cite{TRETYAKOV_Phil.Trans.R.Soc.A_metasurfaces_2015, DING_Rep.Prog.Phys._gradient_2017}, the field of metasurfaces has definitely gained an unprecedentedly strong impetus from the work by the group led by F. Capasso~\cite{Yu2011_science}, which actually ignited the explosive growth afterwards. At the same time, to some extent shaped by this original seminal work, conventional designs of metasurfaces rely predominantly on gradient structures, which can induce spatially nonuniform phase distributions that are essential for many metasurface-based applications. Moreover, a further incorporation of the geometric phase of light into metasurfaces renders extra flexibilities for phase control, which has made accessible lots of advanced functionalities~\cite{WILCZEK_geometric_1989,lin_dielectric_2014,BLIOKH_Nat.Photonics_spinorbit_2015,MAGUID_LightSciAppl._multifunctional_2017}.

Nevertheless, up to now realistic demonstrations of gradient metasurfaces more or less inevitably require the fine discretizations of the structure to generate graded phase profiles~\cite{estakhri_recent_2016,DING_Rep.Prog.Phys._gradient_2017}. Consequently, the fabrication resolution of currently available facilities has imposed a stringent constraint on optical wave control efficiencies based on gradient metasurfaces, especially on those that require large-angle beam routing. Recently, it was demonstrated that beam manipulation  can be made more efficient through proper engineering of the corresponding impedance profiles~\cite{PFEIFFER_metamaterial_2013,TRETYAKOV_Phil.Trans.R.Soc.A_metasurfaces_2015,ESTAKHRI_Phys.Rev.X_wavefront_2016,DIAZ-RUBIO_Sci.Adv._generalized_2017} and/or through geometric optimizations of the employed gradient configurations~\cite{EPSTEIN_Phys.Rev.Lett._synthesis_2016,YANG_Opt.Lett.OL_topologyoptimized_2017,lin2017topology}. However, it further complicates the structures, the functionalities of which are consequently more demanding for fabrication resolutions.  An alternative approach for further improvement is to employ high-index dielectric particles in metasurface, where the resonant excitations and interferences of electric and magnetic multipoles can render enormous extra freedom for magnitude and phase control of optical waves~\cite{jahani_alldielectric_2016,KUZNETSOV_Science_optically_2016,DECKER_J.Opt._resonant_2016,LIU_ArXivPrepr.ArXiv160901099_multipolar_2016}. But, when made graded, the high-index dielectric metasurfaces  still cannot fully solve the problems of relatively low-efficiency of large-angle beam routing and the severe fabrication resolution dependence.

To enable a completely flexible beam manipulation that is simultaneously  highly efficient and  less demanding for advanced fabrication technologies, recently several groups have exploited the grating configurations to take advantage of, rather than to get rid of, the diffraction effect, where the interactions between multipolar interferences  and diffractions play a central role~\cite{WU_NanoLett._experimental_2015,LIN_Sci.Rep._optical_2017,PANIAGUA-DOMINGUEZ_ArXiv170500895Phys._metalens_2017,RADI_Phys.Rev.Lett._metagratings_2017,KHAIDAROV_NanoLett._asymmetric_2017,
CHALABI_Phys.Rev.B_efficient_2017,babicheva_resonant_2017}.
Actually, similar ideas existed in much earlier studies with gratings, where, unfortunately, the multipolar contributions have not been clearly identified or, even, incorrectly attributed to~\cite{CHANG-HASNAIN_Adv.Opt.Photon.AOP_highcontrast_2012,COLLIN_Rep.Prog.Phys._nanostructure_2014,DU_Phys.Rev.Lett._optical_2011}, as we will show below. These work represent a novel trend and clearly indicate that combinations of multipolar interferences and diffractions (or more generally lattice couplings) could provide a new platform for the further development of many branches of nanophotonics related to wavefront engineering. However, to make further progress, there are still  obstacles to surmount: (i)  Most of the existing work have not established the comprehensive connections between multipolar interferences and beam controls , focusing only on the dipolar excitations and interferences only, with the roles of higher-order multipoles neglected~\cite{CHEN_Rep.Prog.Phys._review_2016,DING_Rep.Prog.Phys._gradient_2017,jahani_alldielectric_2016,KUZNETSOV_Science_optically_2016,DECKER_J.Opt._resonant_2016,
DU_Phys.Rev.Lett._optical_2011,RADI_Phys.Rev.Lett._metagratings_2017,CHALABI_Phys.Rev.B_efficient_2017,babicheva_resonant_2017}; (ii) In some studied configurations, the higher-order multipoles could be present and play an indispensable role, which have not been unambiguously identified or studied so far~\cite{CHANG-HASNAIN_Adv.Opt.Photon.AOP_highcontrast_2012,COLLIN_Rep.Prog.Phys._nanostructure_2014,LIN_Sci.Rep._optical_2017,PANIAGUA-DOMINGUEZ_ArXiv170500895Phys._metalens_2017,KHAIDAROV_NanoLett._asymmetric_2017}; (iii) Due to the coherent lattice couplings, the magnitudes and phases of the multipoles excited within each unit cell of a lattice could be significantly different from those of an isolated particle (or particle cluster) that makes the unit cell. Nevertheless, the studies of effects of lattice couplings on multipolar excitations and interferences have been sketchy, especially when higher-order multipoles are present. Only by overcoming such obstacles one can take the full advantages of joint effects of lattice couplings and multipolar interferences to render a new strong impetus for the further developments of related fields of nanophotonics.

In this work, we  analytically study  wave scattering by  one-dimensional (1D) arrays of  high-index dielectric cylinders. Both diffractionless regime of metasurfaces with subwavelength unit cells and diffractive metagratings with unit cell of sizes comparable or larger than the incident wavelength are investigated. Here we have considered interferences between not only the electric and magnetic dipoles, but also their higher-order counterparts. In such  single-layered metalattices, we  have obtained fine controls of optical waves, including perfect transmission, perfect reflection, and large-angle beam steering with high efficiencies. Through detailed and comprehensive analysis on the effects of lattice couplings on the multipolar excitation efficiencies of each cylinder, we reveal that all  demonstrated beam manipulations can be attributed to the interplays between lattice couplings and multipolar interferences, where the induced highly asymmetric angular scattering patterns of each unit cell play an essential role. Basically, such a simple 1D metalattice allows to tackle all the aforementioned problems, and to establish a much broader and more inclusive framework for  multipolar-interference-induced controls of optical wave propagations, which can potentially incubate lots of extra metasurface and metagrating related fundamental studies and applications.

\section{Results and Discussions}

\subsection{Theory for light propagation through 1D metalattices}

\pict[0.8]{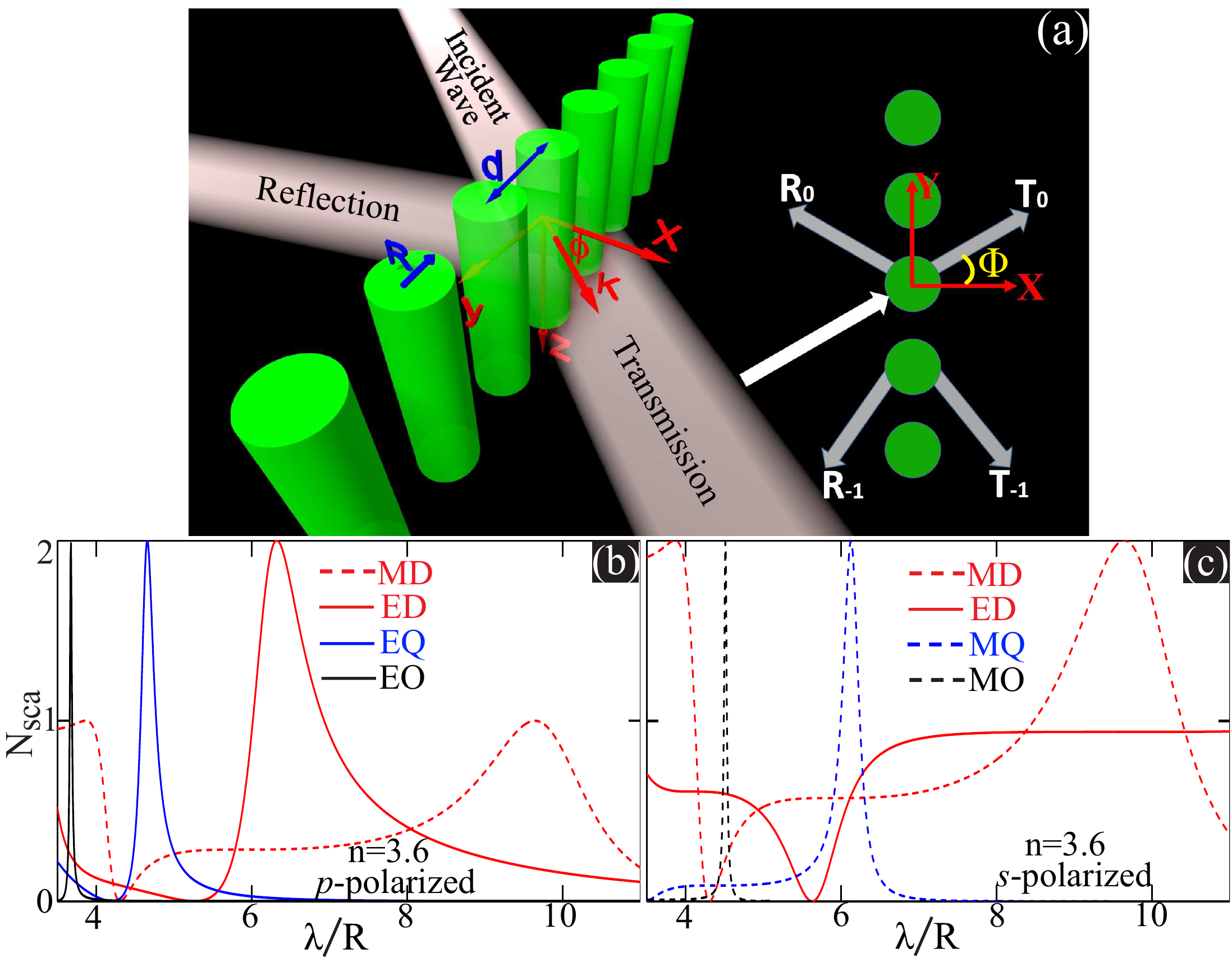}{fig1}{\small (a) Schematic illustration of an normally incident ($\textbf{k} \bot \textbf{z}$) plane wave scattered by a 1D metalattice consisting of dielectric cylinders of refractive index $n$ (set as $n=3.6$ throughout this work), radius $R$ and periodicity $d$. The angle of incidence is $\Phi$ and the wave can be $p$- or $s$-polarized. The metalattice can function as a metasurface without any extra diffractions (main panel) or as a metagrating with emergent diffractions (see inset; $R_{-1}$ and $T_{-1}$ denote respectively reflection and transmission of order $-1$).  (b) and (c) show the normalized scattering spectra contributed separately by the electric and magnetic multipoles (up to octupoles) excited with $p$- and $s$-polarized waves, respectively. The solid and dashed curves correspond to multipoles of electric and magnetic natures, respectively.}

In Fig.~\rpict{fig1}(a) we show the schematic of the 1D metalattice with incident light illumination. It consists of a periodic array of cylinders (radius $R$, and refractive index $n$ which is set as $n=3.6$ throughout this work) with period $d$ along $\textbf{y}$ direction. The  incident plane wave with in-plane wavevector $\textbf{k}$ (in the $\textbf{x}-\textbf{y}$ plane: $\textbf{k} \bot \textbf{z}$) can be $s$-polarized (electric field along $z$ direction: $\mathbf{E_0}\parallel\mathbf{z}$) or $p$-polarized (magnetic field along $z$ direction:  $\mathbf{H_0}\parallel\mathbf{z}$), with incident angle $\Phi$ and free-space wavelength $\lambda$. We investigate the regimes of both diffractionless metasurfaces [with only zeroth order transmission and reflection as shown in Fig.~\rpict{fig1}(a); $\Phi<\arcsin(\lambda/d-1)$] and metagratings [with emergence of higher diffraction  orders $-1$, shown as the inset of Fig.~\rpict{fig1}(a); $\Phi\geq\arcsin(\lambda/d -1)$]. We name the structure under consideration as "metalattices", which is broader term and includes both regimes of metagratings and metasurfaces.

Firstly, we analyse the scattering properties of a single cylinder (radius $R$ and index $n$) under normally incident plane waves. The scattered fields of an isolated cylinder can be expanded into a set of cylindrical harmonics, with the expansion coefficients of $a_m^{p,s}$ for $p$-polarized and $s$-polarized incident waves, respectively~\cite{Bohren1983_book}. Due to the mirror symmetry, there is  a relation between scattering coefficients of opposite orders: $a_m^{p,s}=a_{-m}^{p,s}$. Also, there is an additional constraint  $|a_m^{p,s}|\leq 1$, which is required by the energy conservation. The cylindrical harmonics can be effectively mapped to electromagnetic multipoles~\cite{Bohren1983_book,kallos2012resonance,LIU_Phys.Rev.A_superscattering_2017,liu_generalized_2017}: For $p$-polarization, $a_0^{p}$ corresponds to the magnetic dipole (MD), and $a_{1:3}^{p}$ (or $a_{-1:-3}^{p}$) correspond to the electric dipole (ED), electric quadrupole (EQ), and electric octupole (EO) respectively; While for $s$-polarization, $a_0^{s}$ corresponds to the ED, and $a_{1:3}^{s}$ (or $a_{-1:-3}^{s}$) correspond to the MD, magnetic quadrupole (MQ) and magnetic octupole (MO) respectively. With the expansion coefficients (scattering coefficients) obtained, the scattering cross section of an isolated cylinder can be directly calculated~\cite{Bohren1983_book,kallos2012resonance,LIU_Phys.Rev.A_superscattering_2017,liu_generalized_2017}.  We show the normalized scattering cross section $N_{\rm sca}^{p,s}$ (normalized by the single channel scattering limit $2\lambda/\pi$)  in Figs.~\rpict{fig1}(b) and (c), where only the separate contributes of the excited multipoles are presented.

Based on the multiple scattering theory~\cite{YASUMOTO_ElectromagneticTheoryandApplicationsforPhotonicCrystals_modeling_2005,LIU_ArXivPrepr.ArXiv170406049_scattering_2017,liu_generalized_2017}, the optical properties of the metalattice shown in Fig.~\rpict{fig1}(a) can be obtained through direct analytical calculations. Similar to an isolated cylinder, the scattered fields of the cylinder within the metalattice (lattice-cylinder) can be also expanded into a set of cylindrical harmonics. If  $\breve{a}_{j,m}^{p,s}$ denote the scattering coefficients of the $j^{th}$ lattice-cylinder [with coordinate positions of $y_j=jd$, $x_j=0$ and $z_j=0$ shown in Fig.~\rpict{fig1}(a)], according to the Floquet theory: $\breve{a}_{j,m}^{p,s}$=$\breve{a}_{m}^{p,s}e^{ik_yy_j}$ with $k_y=k\sin\Phi$ ($k=|\textbf{k}|$ is the free-space angular wavenumber).  In sharp contrast to an isolated cylinder, expansion coefficients for the cylinder located at the  origin ($\breve{a}_{m}^{p,s}$) are not constrained by $|\breve{a}_{m}^{p,s}| \leq 1$ any more, and  $\breve{a}_{m}^{p,s}=\breve{a}_{-m}^{p,s}$ is only satisfied for $\Phi=0$, when the mirror symmetry is preserved. To obtain $\breve{a}_{m}^{p,s}$, the lattice coupling should be taken into account, which has actually  incorporated not only the electric-electric multipolar coupling, magnetic-magnetic multipolar coupling, but also the cross coupling between electric and magnetic multipoles of all orders~\cite{LIU_ArXivPrepr.ArXiv170406049_scattering_2017}. Consequently, the scattering coefficients of a lattice-cylinder can be, in general,  different from the isolated counterpart $\breve{a}_{m}^{p,s}\neq {a}_{m}^{p,s}$ (see \textit{e.g.}, Ref.~\cite{Liu2012_PRB}), resulting in completely different angular scattering patterns.  Considering that the optical responses, including transmission, reflection and other diffraction efficiencies, can then be obtained through direct summations of the scattering contributions from all the lattice-cylinders (information contained in $\breve{a}_{m}^{p,s}$ rather than in ${a}_{m}^{p,s}$), it is expected that: (i) The optical feature of a whole structure can not be simply deduced from the scattering of isolated cylinders; (ii)  Higher-order multipoles can directly couple to the fundamental dipoles, and thus the dipolar approximations, then, becomes invalid. At the same time, however, this also means that the roles of higher-order multipoles can render extra flexibilities for metalattice based beam controls; (iii) The lattice couplings can significantly affect the multipolar excitation efficiencies of particles within the lattice, and thus naturally can be employed for beam steering. In other words,  the lattice coupling induced multipolar interferences can result in highly asymmetric angular scattering patterns, which lays the foundation for various sorts of beam manipulations.

\pict[1]{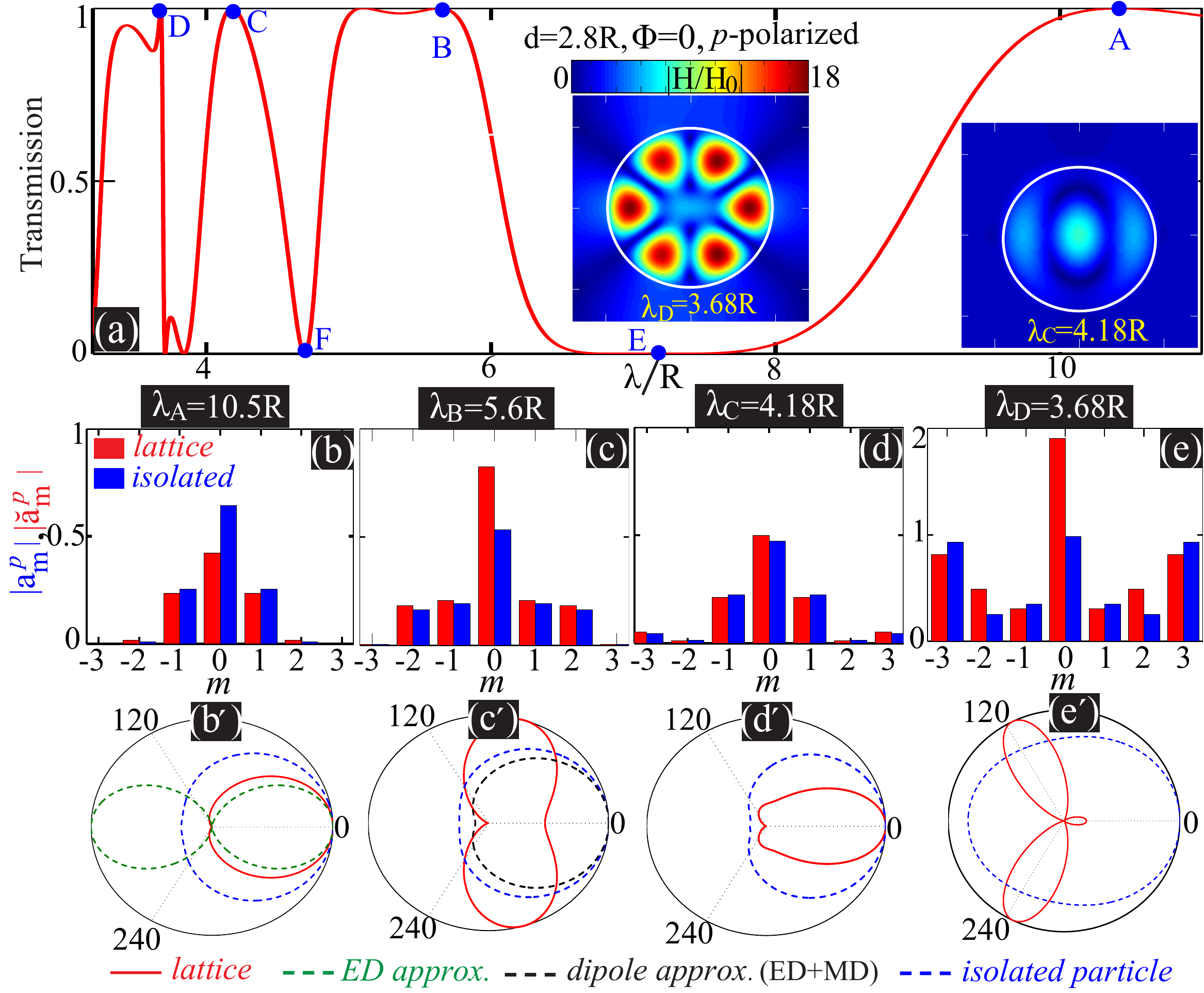}{fig2}{\small (a) Transmission spectra for the metalattice of $d=2.8R$ and $\Phi=0$ with $p$-polarized incident waves. Four perfect transmission points  \textbf{A}-\textbf{D} and two perfect reflection points \textbf{E} and \textbf{F} are pinpointed. The corresponding magnetic near-field distributions at \textbf{C} and \textbf{D} are shown in insets (the white circles denote the cylinder boundaries). (b)-(e) Magnitudes of the multipolar excitation efficiencies (red bar for lattice-cylinder, blue bar for isolated cylinder, which is used throughout this work) for points \textbf{A}-\textbf{D}, respectively. ($\rm b^\prime$)-($\rm e^\prime$) Angular scattering patterns for the lattice-cylinder (solid red curves) and for the isolated cylinder (dashed blue curves) at points \textbf{A}-\textbf{D}, respectively. We also show the results in ($\rm b^\prime$) from the ED approximation at point \textbf{A} (dashed green curve) and the results in ($\rm c^\prime$) from the dipole approximation (only ED and MD are considered: ED+MD) at point \textbf{B} (dashed black curve). }

\subsection{Perfect transmission and reflection in the metasurface regime}
Let's study at the beginning the simplest case of perpendicular incidence ($\Phi=0$) in the diffractionless metasurface regime with only zeroth order transmission and reflection.  The transmission spectra (with a $p$-polarized incident wave and $d=2.8R$) are shown in Fig.~\rpict{fig2}(a), where four points \textbf{A}-\textbf{D} ($\lambda_{\textbf{A}}=10.5R$, $\lambda_{\textbf{B}}=5.6R$, $\lambda_{\textbf{C}}=4.18R$ and $\lambda_{\textbf{D}}=3.68$) of perfect transmission, two points \textbf{E} and \textbf{F} ($\lambda_{\textbf{E}}=7.3R$ and $\lambda_{\textbf{F}}=4.69R$) of perfect reflection are indicated. The magnitudes of the multipolar excitation efficiencies for the four points of complete transparency are summarized in Figs.~\rpict{fig2}(b)-(e), where the results of both the isolated individual cylinder ($|{a}_{m}^{p}|$) and of the lattice-cylinder ($|\breve{a}_{m}^{p}|$) are shown. At points \textbf{A} and \textbf{C} only dipoles (ED and MD) are dominantly excited; While at \textbf{B} and \textbf{D}, all multipoles up to quadrupoles and octupoles have to be taken into account, respectively. The sharp Fano profile around point \textbf{D} originates mainly from the interference between the broad MD and narrow EO. Those results are consistent with what  shown in Fig.~\rpict{fig1}(b). It is also obvious that at all the four points the lattice couplings have effectively changed the multipolar excitation ratios.

To further clarify the underlying mechanism behind those transparency points, the corresponding angular scattering patterns of each cylinder within the lattice are shown in Figs.~\rpict{fig2}($\rm b^\prime$)-($\rm e^\prime$) by solid curves, where we have considered all the multipolar contributions. In addition to that, we have also implemented the following approximations (see the dashed curves): (i) neglect the lattice coupling and show the angular scattering patterns of isolated cylinders for points \textbf{A}-\textbf{D}; (ii) consider only the contributions from electric dipoles (ED approximation: $\breve{a}_{m}^{p}=0$ with $|m|\neq1$) at point \textbf{A}; (iii) consider only the ED and MD dipolar contributions (dipole approximation: ED+MD; $\breve{a}_{m}^{p}=0$ with $|m|>1$) at point \textbf{B}. We note that for all cases, the curves for angular scatterings are normalized, as is the case throughout is work. Figures~\rpict{fig2}($\rm b^\prime$)-($\rm e^\prime$) confirm that, only when both the lattice coupling and all multipolar contributions have been taken into account, the backward scattering can be fully eliminated and then perfect transmission can be obtained [see points \textbf{A}-\textbf{D} in Fig.~\rpict{fig2}(a)]. If either the lattice coupling or any multipoles that have been involved are not considered, the remnant backward scattering will produce noneligible reflections, destroying the ideal metalattice transparencies [this will be further illustrated in  Figs.~\rpict{fig3}(a) and (b)].

Though metasurfaces with perfect transmission have been widely studied~\cite{HOLLOWAY_IEEEAntennasPropag.Mag._overview_2012,YU_NatMater_flat_2014,brongersma2014light,estakhri_recent_2016,zhang_advances_2016,glybovski_metasurface_2016,CHEN_Rep.Prog.Phys._review_2016,
GENEVET_Optica_recent_2017,DING_Rep.Prog.Phys._gradient_2017,jahani_alldielectric_2016,KUZNETSOV_Science_optically_2016,
DECKER_J.Opt._resonant_2016,LIU_ArXivPrepr.ArXiv160901099_multipolar_2016}, the problems of the existing studies are obvious: (i) It is not realized how the perfect transmission is related to multipolar-interference-induced backward scattering elimination\cite{LALANNE_Nanotechnology_antireflection_1997,ARBABI_Nat.Photonics_planar_2017}; (ii) Most of the studies are confined to the dipolar approximation, considering only EDs and/or MDs (see \textit{e.g.},  Refs.~\cite{YANG_NatCommun_alldielectric_2014,decker_high-efficiency_2015}); (iii) Even if it is realized the  contributions from higher-order multipoles are noneligible, their specific roles and the interferences with dipoles are not fully clarified (see \textit{e.g.}, Ref.~\cite{KRUK_APLPhotonics_invited_2016}); (iv) Previous investigations have not revealed in detail how the lattice coupling effects affect the multipolar excitation efficiencies when higher order multipoles are involved~\cite{WU_NatCommun_spectrally_2014}. As a result, the complete correspondences between the full multipolar interferences (and especially the highly asymmetric angular scattering pattern induced) and various sorts of beams controls cannot possibly be established. Here in our work, we have simultaneously solved all those problems.

We further show that, though in the far-field at the four points \textbf{A}-\textbf{D} the metalattice is identically transparent, the corresponding near-field distributions and field enhancement can be contrastingly different. This is, actually, quite expected from the multipolar excitation efficiencies shown in Figs.~\rpict{fig2}(b)-(e). Moreover, we show the near-field distributions (in terms of magnetic field $\textbf{H}_z$) at two indicated points \textbf{C} and \textbf{D} in the insets of Fig.~\rpict{fig2}(a), which share the same color-bar. According to Figs.~\rpict{fig2}(d) and (e), higher-order multipoles with higher efficiencies and higher $Q$-factors are excited at \textbf{D} (including all multipoles up to octupoles) than at \textbf{C} (including  dipoles only). As a result, the field enhancement is more significant at \textbf{D}, while negligible at \textbf{C}, which agrees well with the former classifications of nontrivial and trivial transparencies~\cite{LIU_ArXivPrepr.ArXiv170406049_scattering_2017}.

\pict[0.85]{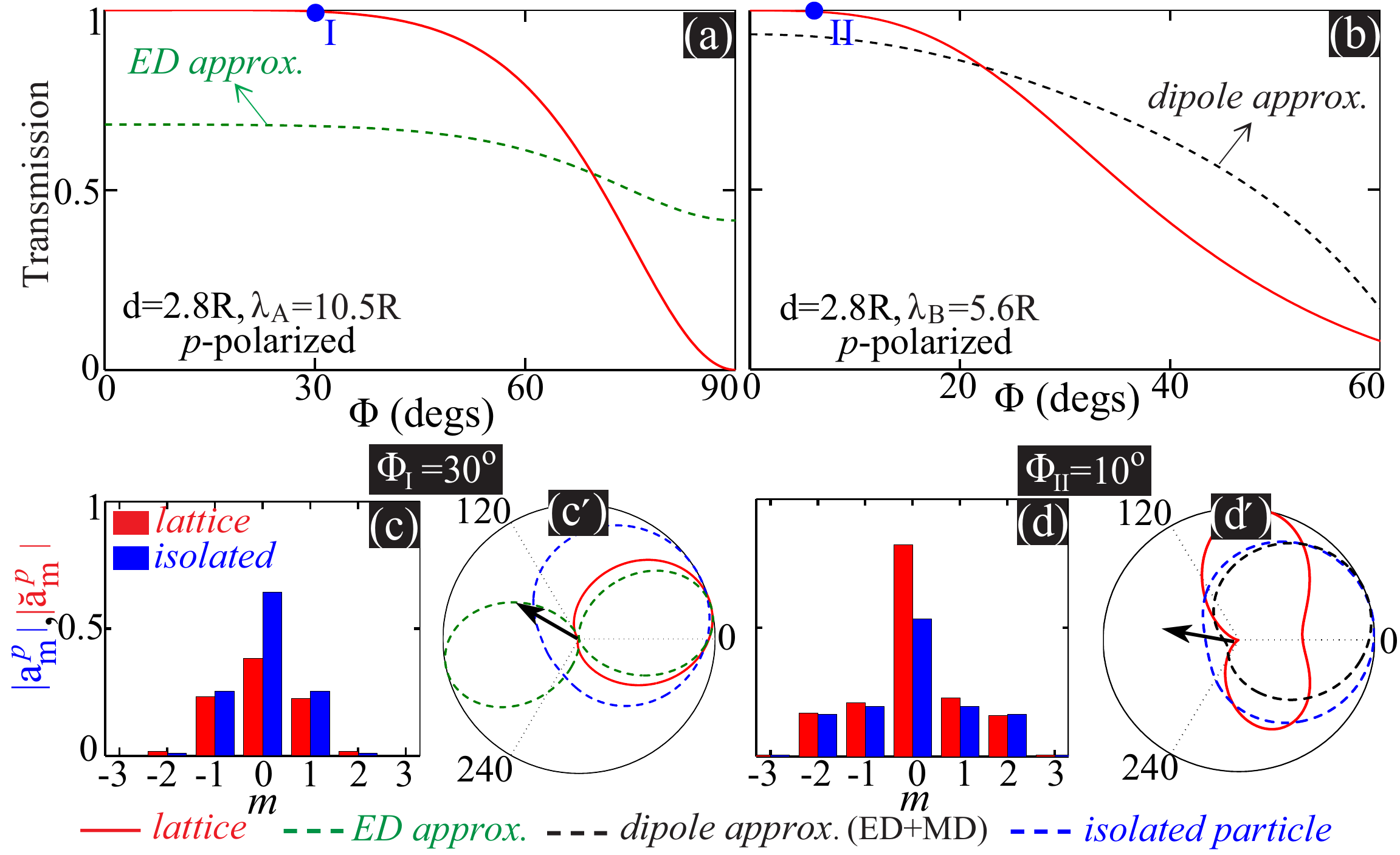}{fig3}{\small  (a) and (b) Transmission spectra with respect to incident angle $\Phi$ at the two points \textbf{A} and \textbf{B} indicated in Fig.~\rpict{fig2}(a). The results from the ED approximation and dipole approximation (ED+MD) are also shown as dashed curves, respectively. Two points of perfect transmission (\textbf{\uppercase\expandafter{\romannumeral1}} and \textbf{\uppercase\expandafter{\romannumeral2}}) with obliquely incident waves are also indicated.  Magnitudes of the multipolar excitation efficiencies at those two points are shown in (c) and (d), respectively. The corresponding angular scattering patterns are shown in ($\rm c^\prime$) and ($\rm d^\prime$), where reflection directions are indicated by  arrows.}

Next, we investigate the dependence of perfect transmission on incident angle $\Phi$ and show the results in Figs.~\rpict{fig3}(a) and (b) for the two points \textbf{A} and \textbf{B} indicated in Fig.~\rpict{fig2}(a). We consider two point
\textbf{\uppercase\expandafter{\romannumeral1}} and \textbf{\uppercase\expandafter{\romannumeral2}} of complete transmission while under oblique incidence ($\Phi_{\textbf{\uppercase\expandafter{\romannumeral1}}}=30^{\circ}$ and $\Phi_{\textbf{\uppercase\expandafter{\romannumeral2}}}=10^{\circ}$) at points \textbf{A} and \textbf{B}, respectively. The corresponding magnitudes of multipolar excitation efficiencies and angular scattering patterns are shown in  Figs.~\rpict{fig3}(c) and ($\rm c^\prime$), and Figs.~\rpict{fig3}(d) and ($\rm d^\prime$), respectively. It is clear that under oblique excitation, the mirror symmetry of the whole structure is broken and thus the relation of $\breve{a}_m^{p,s}=\breve{a}_{-m}^{p,s}$ is violated [more clearly shown in Fig.~\rpict{fig3}(d)]. The preservation of  perfect transmission for oblique incidences originate from the interferences of all the multipoles excited, which results in the elimination of scatterings in the reflection directions [indicated by arrows in Figs.~\rpict{fig3}($\rm c^\prime$) and ($\rm d^\prime$)].  Without considering the lattice coupling or neglecting the contributions from  MD (at point \textbf{\uppercase\expandafter{\romannumeral1}}) or EQ (at point \textbf{\uppercase\expandafter{\romannumeral2}}), the scatterings along the reflection directions will not be fully suppressed [Figs.~\rpict{fig3}($\rm c^\prime$) and ($\rm d^\prime$)] and, thus, the features of full transmission will not be realized anymore [see Figs.~\rpict{fig3}(a) and (b)]. It is worth nothing that here we have, actually, achieved precisely the generalized Brewster angles that were previously demonstrated with coupled dielectric spheres in the dipolar regimes~\cite{PANIAGUA-DOMINGUEZ_NatCommun_generalized_2016-1}. Nevertheless, in this work we provide a more complete and detailed picture, revealing that higher-order multipoles can be also employed and thus render much more flexibilities for beam manipulations.

Let's revisit the two points of \textbf{E} and \textbf{F} indicated in Fig.~\rpict{fig2}(a), where the perfect reflection has been achieved. The dependence of the reflection on incident angle $\Phi$ at those points are shown respectively in Figs.~\rpict{fig4}(a) and (b). At the first glance, according to Fig.~\rpict{fig1}(b) and Fig.~\rpict{fig2}(a), the perfect reflection mainly originate from significant ED (at point \textbf{E}) and EQ (at point \textbf{F}) excitations, respectively.  Nevertheless, detailed results of multipolar excitation efficiencies for the two points of perfect reflection \textbf{\uppercase\expandafter{\romannumeral3}} ($\Phi_{\textbf{\uppercase\expandafter{\romannumeral3}}}=15^{\circ}$) and \textbf{\uppercase\expandafter{\romannumeral4}} ($\Phi_{\textbf{\uppercase\expandafter{\romannumeral4}}}=10^{\circ}$) pinpointed in Fig.~\rpict{fig4}(a) indicate that the contributions of other multipoles are also noneligible [at \textbf{\uppercase\expandafter{\romannumeral3}} the MD should be considered according to Fig.~\rpict{fig4}(c); while at \textbf{\uppercase\expandafter{\romannumeral4}}, besides EQs, both the MD and EDs should be considered according to Fig.~\rpict{fig4}(d)]. This agrees well with the results of reflectively obtained through ED approximation or EQ approximation (considering only the contributions from electric quadrupoles: $\breve{a}_{m}^{p}=0$ with $|m|\neq2$), shown respectively in Figs.~\rpict{fig4}(a) and (b) as dashed curves.

\pict[0.5]{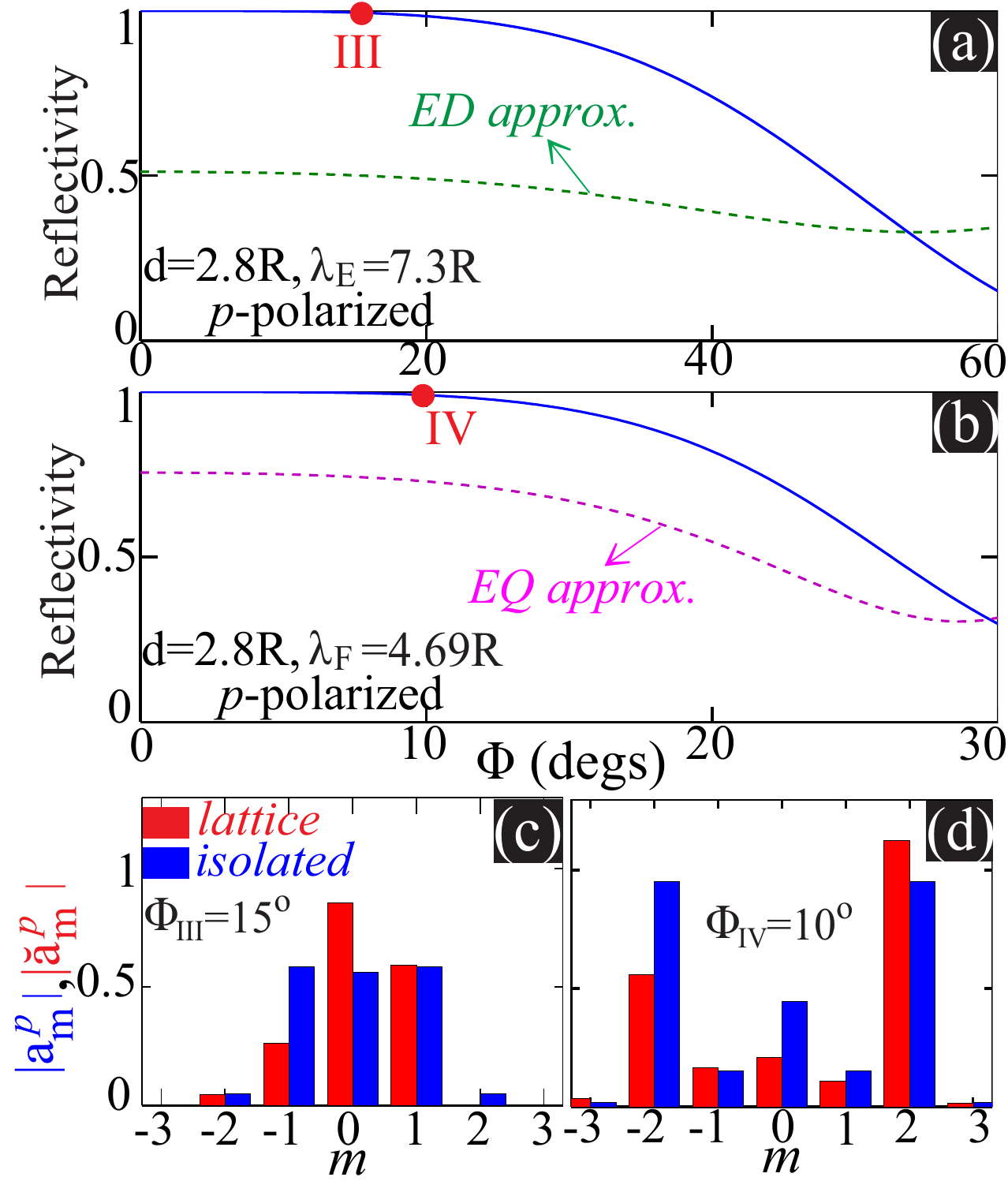}{fig4}{\small (a) and (b) Transmission spectra with respect to incident angle  $\Phi$ at the two points \textbf{E} and \textbf{F} indicated in Fig.~\rpict{fig2}(a). The results from the ED approximation and EQ approximation are also shown as dashed curves, respectively. Two points of perfect transmission (\textbf{\uppercase\expandafter{\romannumeral3}} and \textbf{\uppercase\expandafter{\romannumeral4}}) with obliquely incident waves are also indicated.  Magnitudes of the multipolar excitation efficiencies at those two points are shown respectively in (c) and (d).}

Similar to perfect transmission, the case of perfect reflection has also been widely investigated in lots of metasurface configurations~\cite{HOLLOWAY_IEEEAntennasPropag.Mag._overview_2012,YU_NatMater_flat_2014,brongersma2014light,estakhri_recent_2016,zhang_advances_2016,glybovski_metasurface_2016,CHEN_Rep.Prog.Phys._review_2016,
GENEVET_Optica_recent_2017,DING_Rep.Prog.Phys._gradient_2017,jahani_alldielectric_2016,KUZNETSOV_Science_optically_2016,
DECKER_J.Opt._resonant_2016,LIU_ArXivPrepr.ArXiv160901099_multipolar_2016}. Compared to our comprehensive and thorough analysis, the limitations of former studies include: (i) No complete multipolar investigations have been conducted to reveal the subtle connections between multipolar interferences (especially the induced enhanced  scatterings along the reflection directions) and perfect reflection (see \textit{e.g.}, Refs.~\cite{MATEUS_IEEEPhotonicsTechnol.Lett._ultrabroadband_2004,FATTAL_NatPhoton_flat_2010,CHANG-HASNAIN_Adv.Opt.Photon.AOP_highcontrast_2012,MAGNUSSON_Opt.Lett.OL_wideband_2014,COLLIN_Rep.Prog.Phys._nanostructure_2014,ESFANDYARPOUR_Nat.Nanotechnol._metamaterial_2014,ARBABI_Nat.Photonics_planar_2017});
(ii) Related studies are confined to the dipolar regimes neglecting the roles of higher-order multipoles (see \textit{e.g.}, Refs.~\cite{LIU_OpticaOPTICA_optical_2014,MOITRA_ACSPhotonics_largescale_2015,shalaev_high-efficiency_2015}); (iii) Or higher-order multipolar regimes are investigated, while without clarifying the interferences between all the multipoles involved~\cite{liu_generalized_2017}. We would like to emphasize the work of Ref.~\cite{DU_Phys.Rev.Lett._nearly_2013}, where both the configuration employed and the results obtained are quite similar to ours.  Besides the fact that they have investigated the dipolar regimes only while we here have also considered the regimes of efficient higher multipolar excitations, a more serious problem of Ref.~\cite{DU_Phys.Rev.Lett._nearly_2013} is that the authors incorrectly attribute the effects of perfect reflection to the ED excitation only (rotating electric dipoles with $\breve{a}_1^{p,s}\neq\breve{a}_{-1}^{p,s}$~\cite{DU_Phys.Rev.Lett._nearly_2013}). Actually, this mistake originated from an earlier related work from the same group~\cite{DU_Phys.Rev.Lett._optical_2011}, for which we will provide full clarifications below in Figs.~\rpict{fig6} (c)-($\rm c^\prime$). As has been demonstrated explicitly in Fig.~\rpict{fig4}(c), there are simultaneously efficient MD excitations;  and in  Fig.~\rpict{fig4}(a), by neglecting the contributions from MD the correct results of perfect reflectivity will not be obtained.

Up to now, we have obtained both perfect transmission and perfect reflection with incident $p$-polarized waves, which we will show are also achievable for $s$-polarized waves. Figure~\rpict{fig5}(a) shows the transmission spectra for the metalattice of $d=3.6R$ with perpendicularly incident $s$-polarized waves,  where we have pinpointed two points of perfect reflectivity \textbf{G} and \textbf{K} ($\lambda_{\textbf{G}}=4.525R$ and $\lambda_{\textbf{K}}=6.265R$) and the other two points of perfect transmission \textbf{H} and \textbf{J} ($\lambda_{\textbf{H}}=4.555R$ and $\lambda_{\textbf{J}}=5.864R$). According to Fig.~\rpict{fig1}(c), the Fano profile associated with points \textbf{G} and \textbf{H} is due to the interference of broad ED and narrow MO. The angular scattering patterns  at \textbf{G} and \textbf{H} are shown in insets of Fig.~\rpict{fig5}(a). Similar to the analysis of Figs.~\rpict{fig2}-\rpict{fig4}, the angular scattering patterns agree well with the transmission spectra, where eliminating backward scattering leads to perfect transmission (at \textbf{H}) and perfect reflection is accompanied by significant backward scattering (at \textbf{G}).  The angular dependence of the reflection spectra (at \textbf{G} and  \textbf{K}) and transmission spectra (at \textbf{H} and  \textbf{J}) are shown in Figs.~\rpict{fig5}(b)-(e). For every case an angle is indicated by an star ($\Phi_{\textbf{G}}=14^{\circ}$, $\Phi_{\textbf{K}}=12^{\circ}$, $\Phi_{\textbf{H}}=15^{\circ}$ and $\Phi_{\textbf{J}}=32^{\circ}$) and the corresponding angular scattering patterns (with the reflection directions indicated by the arrows) are shown as the insets. Similar to the perpendicular incident case shown in Fig.~\rpict{fig5}(a) and the insets, at points \textbf{G}-\textbf{K} with obliquely incident waves, the spectra and angular scattering patterns (more specifically zero or nonzero scattering at the reflection directions with the perfect transmission or reflection) agree with each other very well.

\pict[0.5]{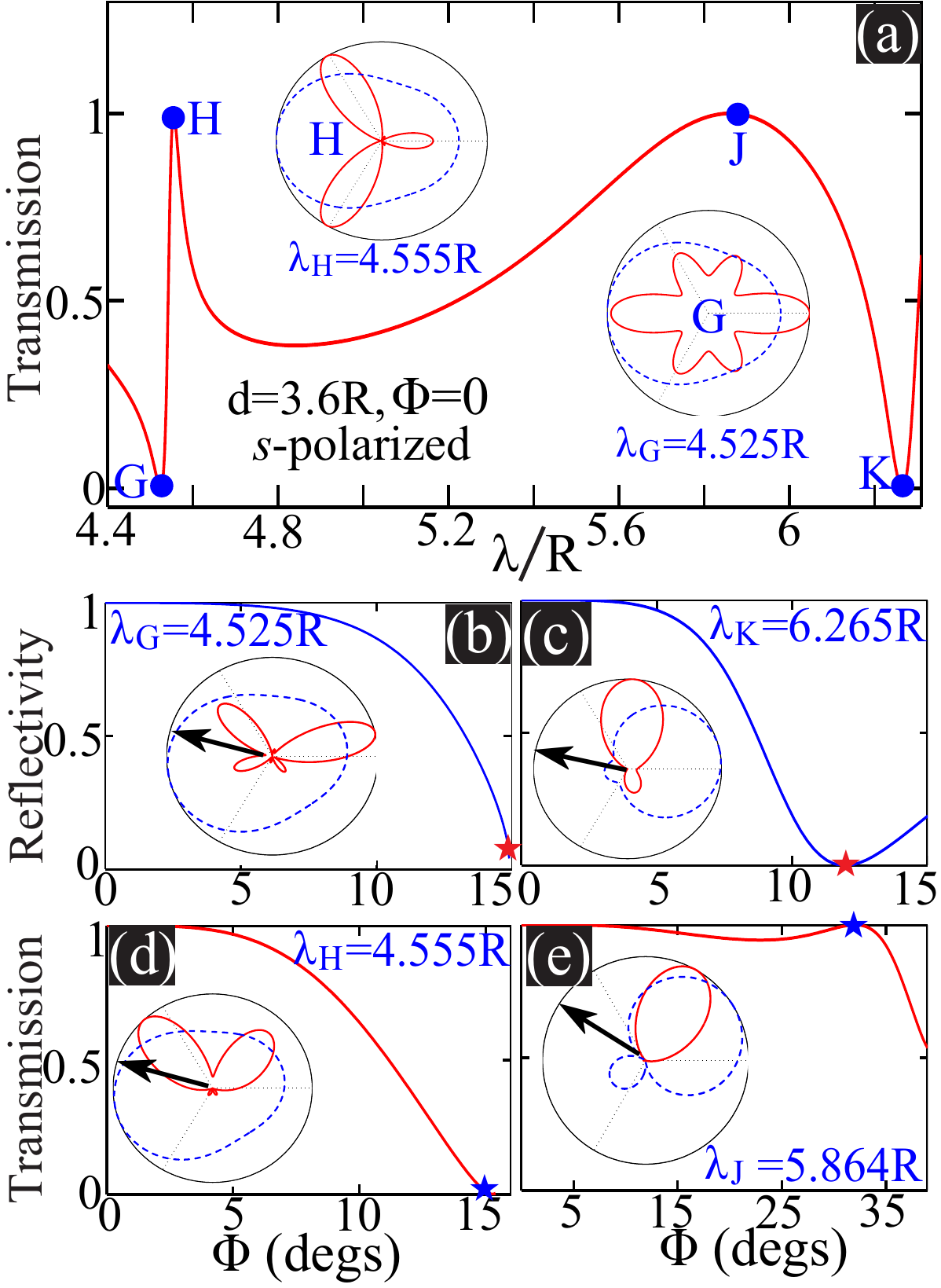}{fig5}{\small (a) Transmission spectra for a metalattice with $d=3.6R$ and $\Phi=0$ with $s$-polarized incident waves. Two perfect transmission points  \textbf{H} and \textbf{J}, and two perfect reflection points \textbf{G} and \textbf{K} are pinpointed. The corresponding angular scattering patterns [solid for lattice-cylinder, dashed for isolated cylinder, as is the case in (b)-(e)] at \textbf{G} and \textbf{H} are shown as insets. (b)-(e) Angular dependence spectra of reflection  (at \textbf{G} and  \textbf{K}) and the transmission (at \textbf{H} and  \textbf{J}). For each point an angle is stared, with the corresponding angular scattering pattern shown as the inset (reflection directions are also indicated by the arrows).}

\subsection{Diffraction management in the metagrating regime}

Now we switch to the metagrating regimes with $\Phi\geq\Phi_c=\arcsin(\lambda/d-1)$ (here $\Phi_c$ is the critical angle for the emergence of diffraction of higher orders beyond zeroth order transmission and reflection), where at least reflection and transmission of order $-1$ will arise [shown as the inset of Fig.~\rpict{fig1}(a)]. Compared to the metasurface regime without extra higher-order diffractions, now it is more challenging to manipulate the optical waves since in the metagrating regimes there are  at least four rather than two out-coupling channels.  We define the polar angle $\theta\in(-\pi,\pi]$ on the $\textbf{x}-\textbf{y}$ plane [see the inset of Fig.~\rpict{fig1}(a)], and then it is easy to obtain the corresponding polar angles for all the four channels as follows: $\theta_{T_0}=\Phi$, $\theta_{R_0}=\pi-\Phi$, $\theta_{T_{-1}}=-\arcsin(\lambda/d-\sin\Phi)$ and $\theta_{R_{-1}}=\arcsin(\lambda/d-\sin\Phi)-\pi$.

We start with $p$-polarized incident waves and show the diffraction spectra with respect to incident angle $\Phi$ in Fig.~\rpict{fig6}(a) for metalattices of $d=4.4R$ and $\lambda=6.27R$, and in Fig.~\rpict{fig6}(b) of $d=2.8R$ and $\lambda=3.71R$.
In Fig.~\rpict{fig6}(a) we indicate a point \textbf{\uppercase\expandafter{\romannumeral5}} ($\Phi_{\textbf{\uppercase\expandafter{\romannumeral5}}}=45^{\circ}$, $\theta_{T_{-1}}\approx-46^{\circ}$ and $\theta_{R_{-1}}\approx-134^{\circ}$) at which the zeroth order transmission and reflection can be totally suppressed, leaving only the reflection and transmission of order $-1$. This means that at point \textbf{\uppercase\expandafter{\romannumeral5}}, close to $80\%$ of the incident wave [see Fig.~\rpict{fig6}(a)] has been bent by approximately $91^{\circ}$ ($\theta_{T_{0}}-\theta_{T_{-1}}$). We note that for gradient metasurfaces it is rather challenging to obtain simultaneously such large bending angle and high efficiency~\cite{estakhri_recent_2016,DING_Rep.Prog.Phys._gradient_2017}. Moreover, according to Fig.~\rpict{fig6}(a), such a functionality can be preserved within a relatively wide incident angle range around $\Phi_{\textbf{\uppercase\expandafter{\romannumeral5}}}=45^{\circ}$.  In sharp contrast, at the indicated point \textbf{\uppercase\expandafter{\romannumeral6}} ($\Phi_{\textbf{\uppercase\expandafter{\romannumeral6}}}=45^{\circ}$, $\theta_{T_{-1}}\approx-38^{\circ}$ and $\theta_{R_{-1}}\approx-142^{\circ}$) in  Fig.~\rpict{fig6}(b), the reflection and transmission orders of  $-1$ have been totally suppressed and the incident waves are split into zeroth order transmission and reflection, rendering the metagrating effectively diffractionless like a metasurface. This difference can be simply explained through investigating the corresponding multipolar excitation efficiencies and the angular scattering patterns at the points selected, which are shown respectively in Figs.~\rpict{fig6}(c) and ($\rm c^\prime$), and Figs.~\rpict{fig6}(d) and ($\rm d^\prime$).  At point \textbf{\uppercase\expandafter{\romannumeral5}}, ED and MD are dominantly excited [see Fig.~\rpict{fig6}(c)] and the interference leads to the suppression of $T_0$ and $R_0$ [see Fig.~\rpict{fig6}($\rm c^\prime$); and  according to the optical theorem~\cite{Bohren1983_book}, the suppression of $T_0$ is induced by the destructive interference between the forward scattering and the incident wave]; while at point \textbf{\uppercase\expandafter{\romannumeral6}}, multipoles up to EO are all effectively excited [see Fig.~\rpict{fig6}(d)], and the full multipolar interferences result in the suppression of $T_{-1}$ and $R_{-1}$ [see Fig.~\rpict{fig6}($\rm d^\prime$)].

 We would like to note that in Ref.~\cite{DU_Phys.Rev.Lett._optical_2011},  similar results of beam steering  to those shown in Fig.~\rpict{fig6}(a) have been obtained. Nevertheless, this by no means compromises the significance of our work, as clearly in Ref.~\cite{DU_Phys.Rev.Lett._optical_2011} a wrong interpretation is provided, attributing the capability of large-angle high-efficiency beam steering to excitations of sole EDs. As we have shown here, without considering the contributions of MD and its interferences with EDs, the reflection can not be effectively suppressed [see the dashed curves in Figs.~\rpict{fig6}(a) and ($\rm c^\prime$)], which would significantly reduce the beam steering efficiencies. Moreover, here we extend the investigations from dipolar regimes to higher-order multipolar regimes with more flexibilities for wave-front controls.

\pict[0.85]{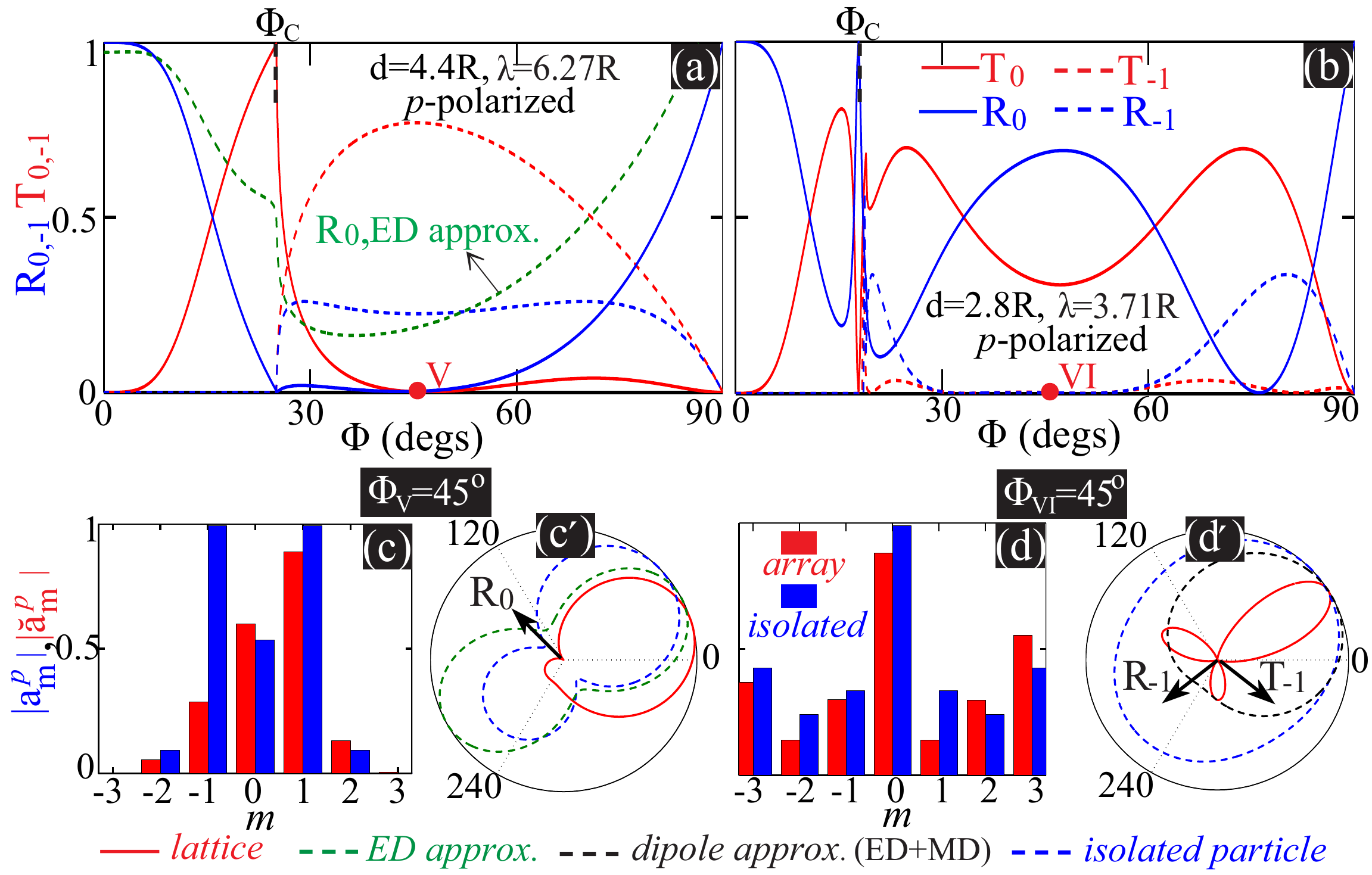}{fig6}{\small Diffraction spectra with respect to incident angle  $\Phi$ for metalattices of  $d=4.4R$, $\lambda=6.27R$ in (a), and of $d=2.8R$, $\lambda=3.71R$ in (b) with incident $p$-polarized waves. In (a) we also show  the reflection spectrum $R_0$ by a dashed green curve that is obtained through the ED approximation. Two points \textbf{\uppercase\expandafter{\romannumeral5}} and \textbf{\uppercase\expandafter{\romannumeral6}} with $\Phi=45^{\circ}$ are indicated. Magnitudes of the multipolar excitation efficiencies at these two points are shown in (c) and (d), respectively. The corresponding angular scattering patterns are shown in ($\rm c^\prime$) and ($\rm d^\prime$), with some of the diffraction directions indicated by arrows. }

\pict[0.85]{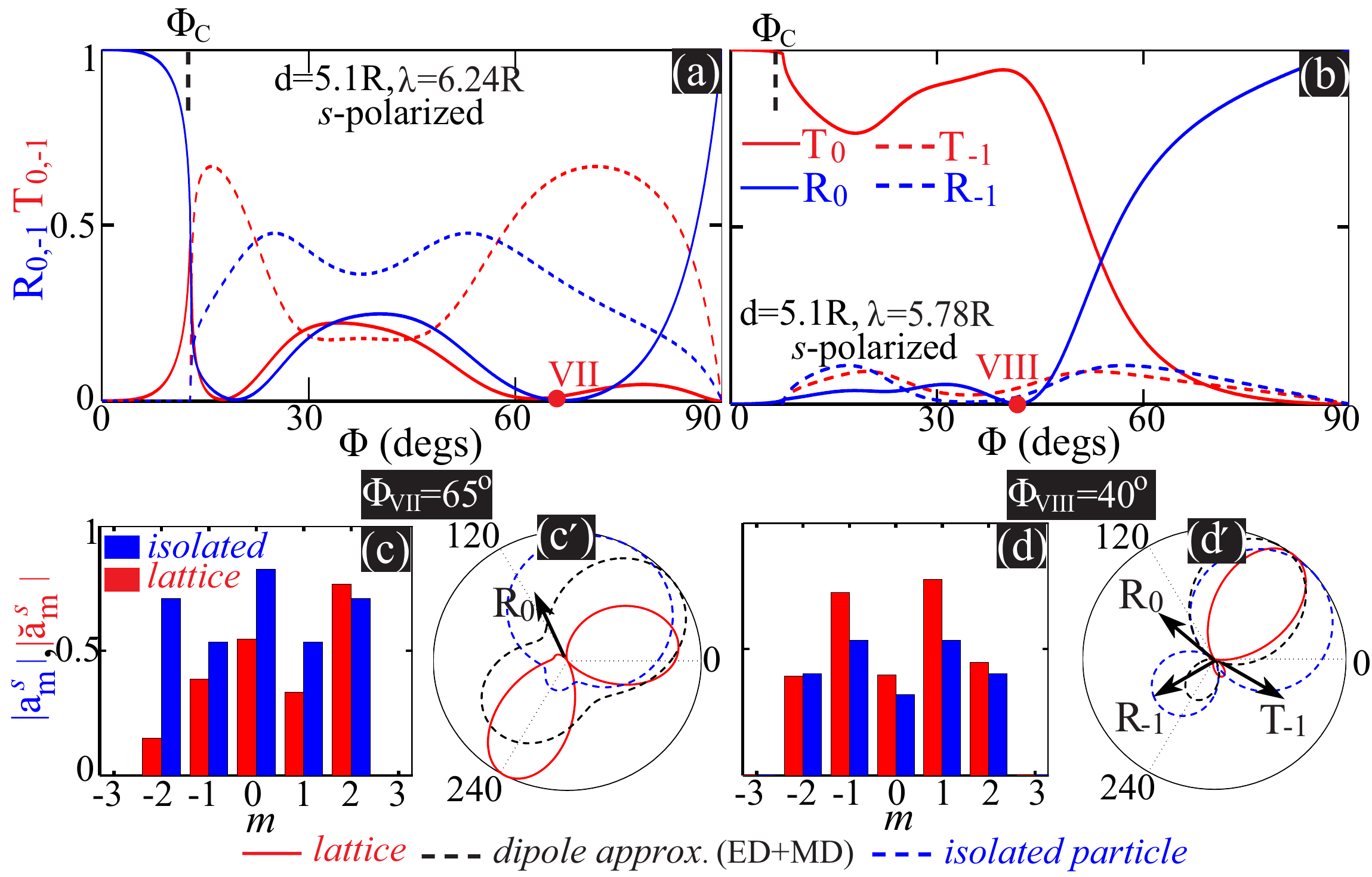}{fig7}{\small Diffraction spectra with respect to incident angle  $\Phi$ for metalattices of $d=5.1R$, $\lambda=6.24R$ in (a), and of $d=5.1R$,$\lambda=5.78R$ in (b) with incident $s$-polarized waves.  Two points \textbf{\uppercase\expandafter{\romannumeral7}} and \textbf{\uppercase\expandafter{\romannumeral8}} are indicated. Magnitudes of the multipolar excitation efficiencies at these two points are shown in (c) and (d), respectively. The corresponding angular scattering patterns are shown in ($\rm c^\prime$) and ($\rm d^\prime$), with some of the diffraction directions indicated by arrows.}

In contrast to previous studies that are limited to $p$-polarized incident waves (see \textit{e.g.}, Refs.~\cite{WU_NanoLett._experimental_2015,DU_Phys.Rev.Lett._optical_2011}), we further demonstrate that such large-angle high-efficiency beam steering is also achievable for $s$-polarized waves. The related spectra with $d=5.1R$ and $\lambda=6.24R$ is shown in Fig.~\rpict{fig7}(a), and with $d=5.1R$ and $\lambda=5.78R$ in Fig.~\rpict{fig7}(b).
A point \textbf{\uppercase\expandafter{\romannumeral7}} ($\Phi_{\textbf{\uppercase\expandafter{\romannumeral7}}}=65^{\circ}$, $\theta_{T_{-1}}\approx-18.5^{\circ}$ and $\theta_{R_{-1}}\approx-161.5^{\circ}$) is pinpointed in  Fig.~\rpict{fig7}(a), where the zeroth order transmission and reflection can be totally suppressed, with only the higher-order diffractions being present. For example, close to $70\%$ of the incident wave has been bent by approximately $83.5^{\circ}$ ($\theta_{T_{0}}-\theta_{T_{-1}}$) into the ${T_{-1}}$ direction [see Fig.~\rpict{fig7}(a)]. In contrast, at indicated point \textbf{\uppercase\expandafter{\romannumeral8}} ($\Phi_{\textbf{\uppercase\expandafter{\romannumeral8}}}=40^{\circ}$, $\theta_{T_{-1}}\approx-29^{\circ}$ and $\theta_{R_{-1}}\approx-151^{\circ}$) in  Fig.~\rpict{fig7}(b), not only the diffractions of order $-1$ have been significantly suppressed but also the zeroth order reflection has been totally eliminated. As a result, the metagrating at point \textbf{\uppercase\expandafter{\romannumeral8}} is almost transparent, with transmission close to $95\%$. Compared to the total transparencies shown in Figs.~\rpict{fig2}, ~\rpict{fig3} and ~\rpict{fig5} for metasurfaces, it is more challenging to achieve the same with metagratings, as it is required to significantly suppress or eliminate three out-coupling channels rather than only one zeroth order reflection channel. The corresponding multipolar excitation efficiencies and the angular scattering patterns are shown  in Figs.~\rpict{fig7}(c) and ($\rm c^\prime$), and Figs.~\rpict{fig7}(d) and ($\rm d^\prime$), respectively.  It is clear that the suppressions of reflections [see Figs.~\rpict{fig7}(a) and (b)] and higher-order diffractions [see Fig.~\rpict{fig7} (b)] originate from the multipolar interferences involving ED, MD and MQ [see Figs.~\rpict{fig7}(c) and (d)], which has induced the scattering suppression or elimination at corresponding diffraction directions [see Figs.~\rpict{fig7}($\rm c^\prime$) and ($\rm d^\prime$)].

 Compared to previous studies for large-angle high-efficiency beam steering relying an asymmetric unit cells ~\cite{LIN_Sci.Rep._optical_2017,PANIAGUA-DOMINGUEZ_ArXiv170500895Phys._metalens_2017,KHAIDAROV_NanoLett._asymmetric_2017}, the results in Figs.~\rpict{fig6} and ~\rpict{fig7} clearly demonstrate that asymmetric antennas (like the one shown in Ref.~\cite{ALBELLA_Sci.Rep._switchable_2015} for example) are not really essential. The reason to employ asymmetric configurations is to produce highly asymmetric scattering patters.  In contrast, we have demonstrated that when the interplay between the lattice coupling and full multipolar interferences is carefully engineered, a basic symmetric and homogeneous particle as the each unit cell would be more than sufficient to generate the asymmetric angular scattering patterns that are required for many sorts of metalattice-based beam controls.

\section{Conclusions and Outlook}
In conclusion,  in this work we have shown explicitly through analytical studies that by carefully manipulating the interplays between multipolar interferences and lattice couplings, highly efficient beam steering can be obtained  with rather simple metalattice configurations. Specifically, we investigated light beam scattering through 1D metalattices consisting of high-index dielectric cylinders, covering both the regimes of diffractionless metasurfaces and diffractive metagratings.  It is revealed that the lattice couplings have  strong effect on the multipolar excitation efficiencies (including both fundamental zeroth order and higher order electric and magnetic multipoles) of the lattice-cylinders coupled to one another, which results in highly asymmetric scattering patterns and thus enables efficient beam manipulations including perfect reflection, perfect transmission and large-angle beam steering.

We note that in this present work we have confined our study to the simplest case of 1D metalattices consisting of the fundamental homogeneous cylindrical particles.  Similar investigations can certainly be extended to 2D metalattices made of spheres, and other homogeneous or inhomogeneous particles of irregular shapes
 with possibly more higher-order diffractions. Considering that photonic dielectric structures are fully scalable, the conclusions we draw hold for various spectral regimes. At the same time, the principles revealed can certainly be extended from periodic lattices to quasi-periodic or random lattices. It would thus be even more promising when all the effects including lattice couplings,  multipolar interferences,  photonic bandgaps (periodic structures) or Anderson localization (random structures), topological effects and so on are combined, through integrating metalattices with the emerging families of 2D materials and/or topological structures, which can potentially stimulate more exotic fundamental studies and more advanced and sophisticated applications in both classical and quantum optics regimes.

\section{Methods}
\subsection{Scattering by an isolated cylinder}
For an individual cylinder with a perpendicularly incident plane wave (propagation direction perpendicular to cylinder axis), the scattering cross section normalized by the single channel scattering limit $2\lambda/\pi$ is~\cite{Bohren1983_book}:
\begin{equation}
\label{Q_ext}
N_{\rm sca}^{p,s} = \sum\nolimits_{m = -\infty}^\infty|a^{p,s}_m|^2,
\end{equation}
where $a_m^{p,s}$ are the scattering coefficients for $p$- and $s$-polarized incident waves respectively, which can be calculated analytically as follows~\cite{Bohren1983_book}:
\begin{eqnarray}
\label{scattering coefficients}
a_m^{p}=a_{-m}^{p}={{n\textbf{J} _{ m} (n\alpha )\textbf{J}'_{m} (\alpha ) - \textbf{J}_{m} (\alpha )\textbf{J}'_{m} (n\alpha )} \over {n\textbf{J} _{ m} (n\alpha )\textbf{H} '_{m} (\alpha ) - \textbf{H} _{m} (\alpha )\textbf{J} '_{ m} (n\alpha )}},\\
a_m^{s}=a_{-m}^{s}={{n\textbf{J} _{ m} (\alpha )\textbf{J}'_{m} (n\alpha ) - \textbf{J}_{m} (n\alpha )\textbf{J}'_{m} (\alpha )} \over {n\textbf{J} '_{ m} (n\alpha )\textbf{H} _{m} (\alpha ) - \textbf{H} '_{m} (\alpha )\textbf{J} _{ m} (n\alpha )}}.
\end{eqnarray}
Here $\textbf{J}$ and $\textbf{H}$ are respectively the first-kind Bessel and Hankel functions; $\alpha$ is the normalized size parameter $\alpha=kR$; and the accompanying primes indicate their differentiation with respect to the entire argument.

\subsection{Optical properties of the 1D metalattice}

According to the multiple scattering theory~\cite{YASUMOTO_ElectromagneticTheoryandApplicationsforPhotonicCrystals_modeling_2005}, $\breve{a}_m^{p,s}$ is related to ${a}_m^{p,s}$ through:
\begin{equation}
\label{lattice_equation}
(\widehat{I}-\widehat{T}\cdot\widehat{C})\textbf{A}=\widehat{T}\textbf{B},
\end{equation}
where $\widehat{I}$ is the identity matrix; $T_{ml}=-\delta_{ml}a_m^{p,s}$ ($\delta_{ml}$ is Kronecker delta function); $\textbf{A}=\breve{a}_m^{p,s}$; $\textbf{B}=\{e^{-im\Phi}(i)^m\}$; and the lattice coupling effect is embedded into the lattice sum matrix $C_{ml}$. This matrix can be expressed as~\cite{YASUMOTO_ElectromagneticTheoryandApplicationsforPhotonicCrystals_modeling_2005}:

\begin{equation}
\label{lattice_matrix}
C_{ml}=\sum\nolimits_{j = 1}^\infty  {\textbf{H}_{m - l} (jkd)[e^{ik_yj d}  + ( - 1)^{m - l} e^{ - ik_yj d} ]},
\end{equation}
 which can be more efficiently calculated through its integral form~\cite{YASUMOTO_ElectromagneticTheoryandApplicationsforPhotonicCrystals_modeling_2005}. Through solving Eq.~(\ref{lattice_equation}) with Eq.~(\ref{lattice_matrix}), the field expansion coefficients $\breve{a}_m^{p,s}$ can be obtained. In a similar way, through implementing the boundary conditions, the field expansion coefficients inside the cylinders can be also obtained~\cite{YANG_IEEETrans.Geosci.RemoteSens._twodimensional_2005-1}, and thus the fields both inside and outside of the cylinders can be directly calculated. The angular scattering  pattern (in terms of field intensity) $\breve{\Gamma}(\theta)$ for the lattice-cylinder can be expressed as:

\begin{equation}
\label{angular patterns}
\breve{\Gamma}^{p,s}(\theta )\propto \left|\sum\limits_{m= -\infty}^\infty  {\breve{a}_m^{p,s}\exp [im(\theta-\Phi) ]} \right|^2,
\end{equation}
and the angular scattering patterns for isolated cylinders ${\Gamma}(\theta)$ can be directly obtained by replacing $\breve{a}_m^{p,s}$ with ${a}_m^{p,s}$ in Eq.~(\ref{angular patterns}).

For the whole metalattice, the reflection and transmission of the $v^{th}$ diffraction order can be expressed as~\cite{YASUMOTO_ElectromagneticTheoryandApplicationsforPhotonicCrystals_modeling_2005}:
\begin{eqnarray}
\label{reflection_coefficients}
R^{p,s}_v(\Phi)=|r^{p,s}_v(\Phi)|^2  =\left|\frac{2}{d\sqrt{k_xk_{xv}}}\sum\nolimits_{m = -\infty}^\infty(-\frac{ik_{yv}+k_{xv}}{k})^m\breve{a}_m^{p,s}\right|^2,\\
\label{reflection_coefficients_2}
T^{p,s}_v(\Phi)=|t^{p,s}_v(\Phi)|^2 =\left|\delta_{0v}+\frac{2}{d\sqrt{k_xk_{xv}}}\sum\nolimits_{m = -\infty}^\infty(-\frac{ik_{yv}-k_{xv}}{k})^m\breve{a}_m^{p,s}\right|^2,
\end{eqnarray}
where $v$ is the diffraction order, and $v=0$ corresponds to fundamental zeroth order reflection and transmission [see Fig.~\rpict{fig1}(a) and the inset]; $k_y=k\sin\Phi$ and $k_{yv}=k_y+2v\pi/d$; $k_{xv}=(k^2-k_{yv}^2)^{-1/2}$; The forward  zeroth order transmission consists of both scattered and incident waves, as a result there is an extra term $\delta_{0v}$ for $T^{p,s}_0(\Phi)$. It is worth nothing that Eqs.~(\ref{reflection_coefficients}) and (\ref{reflection_coefficients_2}) are valid only if $k_{xv}$ is real. Otherwise, $R^{p,s}_v, T^{p,s}_v(\Phi)=0$..

\section{Acknowledgements}
We are indebted to Y. S. Kivshar for initial inspiring suggestions and discussions, and acknowledge a financial support from the National Natural Science Foundation of China (Grant number: $11404403$), the Australian Research Council and the Outstanding Young Researcher Programme of the National University of Defense Technology.
\bibliographystyle{achemso}
\providecommand*\mcitethebibliography{\thebibliography}
\csname @ifundefined\endcsname{endmcitethebibliography}
  {\let\endmcitethebibliography\endthebibliography}{}

\end{document}